\documentclass[12pt]{article}
\usepackage{a4,epsf}
\newcommand{\bbox}[1]{\mbox{\boldmath ${{#1}}$}} 
\newcommand{\bdelta}{\bbox{\delta}}
\newcommand{\bnabla}{\bbox{\nabla}}

\begin{document}

%\vspace*{10mm}
\begin{center}
  {\Large \sffamily Disorder Driven Lock-In Transitions of CDWs \\ and
    Related Structures}\\[.5cm]

  Thomas Nattermann, Thorsten Emig and Simon Bogner\\ {\em \small
    Institut f\"ur Theoretische Physik, Universit\"at zu K\"oln,
    Z\"ulpicherstr. 77,\\ D-50973 K\"oln, Germany}\\[.3cm] 
\end{center}

\begin{quote}
{\small {\bf Abstract.} 
  Thermal fluctuations are known to play an important role in
low-dimensional systems which may undergo incommensurate-commensurate
or (for an accidentally commensurate wavevector) lock-in
transitions. In particular, an intermediate floating phase with
algebraically decaying correlations exists only in $D=2$ dimensions,
whereas in higher dimensions most features of the phase diagram are
mean-field like.

Here we will show, that the introduction of frozen-in disorder leads
to strong fluctuation effects even in $D<4$ dimensions.  For
commensurate wavevectors the lattice pinning potential dominates
always over weak impurity pinning if $p \le p_c=6/\pi$ ($D=3$), where $p$
denotes the degeneracy of the commensurate phase.  For larger p a
disorder driven continuous transition between a long-range ordered
locked-in phase and quasi-long-range ordered phase, dominated by
impurity pinning, occurs. Critical exponents of this transition, which
is characterized by a zero temperature fixed point, are calculated
within an expansion in $4-D$. The generalization to incommensurate
wavevectors will be discussed.  If the modulation in the
quasi-long-range ordered phase has hexagonal symmetry, as e.g.for
flux-line lattices, the algebraic decay is non-universal and depends
on the Poisson ratio of the elastic constants. Weakly driven transport
is dominated by thermally activated creep in both phases, but  with
different creep exponents.}
\end{quote}

%{\bf 1. FIRST SECTION}\\

Incommensurate (I) phases appear in a large variety of systems
(for a review see e.g. [1]). 
Examples are:\\
(i) Charge density waves  in quasi one and two
dimensional conductors (e.g. in TTF--TCNQ, 2H--TaSe).\\
(ii) Spin density waves  (e.g. in ${\rm CuGeO}_3$).\\
(iii) Mass density waves in adsorbed monolayers (e.g.
He, Kr on graphite or metal surfaces) or of reconstructed surfaces 
(e.g. of Mo).\\
(iv) Polarization density waves  in ferroelectrics with
an incommensurate phase (e.g. in ${\rm K}_2{\rm SeO}_4$).\\
(v) Flux density waves  in type II superconductors or
Josephson junctions in an external field.

Charge density waves are usually accompanied by a mass density wave 
like in superionic conductors or
reconstructed metal surfaces.
A common feature of the I phase is, that the wave vector ${\bf q}_0$,
which describes the spatial modulation of the density wave in
the absence of any coupling to the lattices, varies continuously
with the parameter of the system (e.g. temperature $T$, pressure $p$,
chemical potential $\mu$, magnetic field $H$ etc.). If $2\pi /q_0$ is 
close to a multiple of the spacing of the underlying crystal lattice, i.e. if
$|{\bf g}/p-{\bf q}_0|=|\bdelta|<\delta_c$,
commensurability effects may become important. Here ${\bf g}$
denotes a reciprocal lattice vector of the crystal and $p$
is an integer. The modulation then may become commensurate (C) with the
crystal lattice. The most striking effect of the C--phase is the existence of
a gap in the excitation spectrum, in contrast to the I--phase, where the
low--lying excitations are gapless.

The systematic {\it mean--field theory} of the IC--transition was worked out by
Bruce, Cowley and Murray in 1978 [2]. These authors distinguish IC--transition
of type--I and type--II, depending on the absence or existence, 
respectively, of an inversion symmetry around ${\bf g}/p$
in the (disordered phase) soft mode dispersion. In the most simple
case of type--I transition, condensation takes place only on wave 
vectors $\pm {\bf q}_0$. For temperatures sufficiently below the 
mean--field transition temperature $T_{c0}$ the system can be 
described, ignoring amplitude fluctuations, by the sinus--Gordon Hamiltonian.
   \begin{equation}
   {\cal H}=\gamma\int {\rm d}^Dx\left\{\frac{1}{2}(\bnabla\phi -\bdelta)
   -g^2v\cos{p\phi}\right\}\,,
   \label{eq:H}
   \end{equation}
where $\phi({\bf x})$ describes the long--wavelength distortions of
the charge (spin, mass, flux etc.) density
   \begin{equation}
   \rho({\bf x})=\rho_0+\rho_1\,{\rm Re}\left\{e^{i\left(
   \frac{1}{p}{\bf g}{\bf x}+\phi({\bf x})\right)}\right\}\,.
   \label{eq:rho}
   \end{equation}
Minimization of (\ref{eq:H}) yields for $\delta <\delta_c\approx gv^{1/2}$
the solution of the C--phase
$\phi=\frac{2\pi}{p}n\;(n=0,1,\ldots ,p-1)$. The solution for
$\delta >\delta_c$ is a regular lattice of phase--solitons of
distance 
$l\sim \left|\ln{\left(\frac{\delta -\delta_c}{\delta_c}\right)}\right|$ and
internal widths $\xi_0\approx 1/(pgv)^{1/2}$, which describe the
I--phase close to the IC--transition. Far from the transition on has  
$\phi({\bf x})\simeq\bdelta {\bf x}$. Type--I transitions are therefore
continuous. In general, a large number of C--phases is possible
which may lead in certain lattice models to a {\it devil's staircase} behaviour
of the modulation vector as a function of the misfit $\delta$ [1]. 

Type--II transitions, on the contrary, are discontinuous and
show in the I--phase an almost sinusoidal modulation of the order parameter. 
A recent example is the spin--Peierls system ${\rm CuGeO}_3$ [3].\bigskip

{\it Thermal fluctuations} have a strong effect on the IC--phase
diagram in $D=2$ dimensions. If we exclude topological defects (i.e.
vortices), then\\ 
(i) the IC transition becomes an inverted
Beresinskii--Kosterlitz--Thouless--transition with a reduced transition
temperature $T_{c}(\delta =0)\approx 8\pi\gamma /p^2$.\\ (ii) Inside
the I--phase solitons interact now by entropic repulsion.  Close to
the IC--transition the soliton--distance $l$ increases as power law
$ l\sim (\delta -\delta_c)^{-\beta_s}$ where 
$\beta_s =\frac{\zeta}{2(1-\zeta)}$.
 Here $\zeta$ is the thermal roughness exponent of a single soliton 
(domain wall) 
$\zeta_{\rm th}=(3-D)/2=1/2$.\\
(iii) The spatial variation of the density 
$\delta\rho({\bf x})=\rho({\bf x})-\rho_0$
shows in the I--phase only quasi long range order (LRO):
   \begin{equation}
   K({\bf x})=\big<e^{i\big(\phi({\bf x})-\phi({\bf 0})\big)}\big>\sim
   |{\bf x}|^{-\eta} cos(2\pi z/pl)\,,
   \label{eq:K}
   \end{equation}
where ${\bdelta}=\delta {\bf e_z}$. $\eta$ depends 
on $l$, $T/\gamma$ and $p$ and 
 approaches $2/p^2$ for $l\to\infty$.\\
(iv) Topological defects diminish further the ordered (C and I) phases and 
even change the topology of the phase diagramm
for $p\le 4$. In particular, for $p\le 2$ there is no
direct IC--transition, both phases are separated by a fluid phase [4].
Closely results are expected for $1-$dimensional quantum systems.
It is interesting to remark, that a qualitatively  similar picture emerges
also in lower dimensions $1<D<2$, but with different singularities at
the transitions [5].

On the contrary, in three (and higher) dimensions, thermal 
fluctuations have only minor effects on the phase diagram and are 
relevant essentially only in the critical region [6].

\bigskip

In the rest of the paper we investigate the influence of
randomly distributed {\it frozen impurities} on the lock--in transition.
Since impurities favour certain values of the phases, the impurity
Hamiltonian can be written in the form
   \begin{eqnarray}
   {\cal H}_{\rm imp} & = & \gamma\int{\rm d}^Dx\,V({\bf x},\phi)\nonumber\\
   V({\bf x},\phi) & = & \sqrt{\Delta}\cos{\big(
   \phi({\bf x})-\alpha({\bf x})\big)}\,,\quad
   \Delta =\rho_1^2V_0^2n_{\rm imp}
   \label{eq:Himp}
   \end{eqnarray}
$\alpha({\bf x})$ is a randomly frozen phase $(0\le\alpha\le 2\pi)$,
$\gamma V_0$ and $n_{\rm imp}$ denote the strength and the concentration,
respectively, of the impurities.
The modell defined by (\ref{eq:H}) and (\ref{eq:Himp}) describes also an 
$XY$--model in a crystalline  (corresponding to a $p$--fold
axis) and a random field.

\bigskip

We will first discuss the case of a vanishing lattice
potential, $v=0$, and exclude topological defects for most parts of the
rest of the paper. Larkin  and in the present context first Fukuyama
and Lee [7] have shown, that the impurities destroy the translational LRO 
of the charge density wave on scales
$L \gg L_{\xi}\approx \Delta^{1/(4-D)}$ in all dimensions
$D\le 4$. Later studies[8] have shown, that for $2<D<4$ $ K({\bf x})$
 decays as a power
   \begin{equation}
   K({\bf x})\sim e^{i\bdelta {\bf x}}
   \left|\frac{{\bf x}}{L}\right|^{-\bar\eta}\,,
   \label{eq:K1}
   \end{equation}
with a universal exponent 
$\bar\eta=\left(\frac{\pi}{3}\right)^2\epsilon$
in lowest order in $\epsilon =(4-D)$. Thus, we regain now quasi--LRO in 
the I-phase in all dimensions $2<D<4$.

In systems in which the modulation of the I--phase has hexagonal
symmetry, like in flux line lattices, the situation is more complicated.
If we describe the distortions of flux lines by the displacement field
${\bf u}({\bf x})$, the relevant correlation function to
describe long range translational order is given by
$K_{\bf G}({\bf x})={\big<e^{i{\bf G}\big({\bf u}({\bf x})-{\bf u}({\bf 0})\big)}\big>}$, where ${\bf G}$ denotes the reciprocal lattice vector of the
Abrikosov lattice. Recently, Emig at al. [9]
found from a functional renormalization group (FRG) calculation
   \begin{equation}
   K_{\bf G}({\bf x})\sim\left[L_{\xi}^{\phantom{1}}
   ({\bf x}_{\perp}^2+\kappa z^2_l)^{-\frac{1}{2(1+\kappa)}}
   ({\bf x}_{\perp}^2+ z^2_l)^{-\frac{1}{2(1+1/\kappa)}}
   \right]^{\bar\eta_{\bf G}}\,,
   \label{eq:KG1}
   \end{equation}
where ${\bf x}=({\bf x}_{\perp},z)$, $z_l=(c_{11}/c_{44})^{1/2}z$ and
$\kappa =c_{66}/c_{11}$. $c_{11},c_{44}$ and $c_{66}$ are the effective
elastic constants of the flux line lattice (renormalized by thermal
and disorder effects). In the marginal cases $\kappa =0$ and $\kappa =1$
one finds from (\ref{eq:KG1}) for the structure factor
    \begin{equation}
    S({\bf G}+{\bf q})=\int{\rm d}^3_{\phantom{1}}x\,
    e^{i{\bf q}{\bf x}}_{\phantom{1}}
    K_{\bf G}^{\phantom{1}}({\bf x})\sim\left({\bf q}^2_{\perp}+
    \frac{c_{44}}{c_{66}}q^2_z
    \right)^{-(3+\bar\eta_{\bf G})/2}_{\phantom{1}}\,,
    \label{eq:SG}
    \end{equation}
i.e. the structure factor exibits {\it Bragg--peaks}. The exponent
$\bar\eta_{\bf G}$ is {\it non--universal} and depends on the
value of $\kappa$ ($1.143\le\bar\eta_{{\bf G}_0}\le 1.159$ for
$1\ge\kappa\ge 0$). In a large range of external fields 
$\kappa\approx\phi_0/16\pi\lambda^2B$ such that one could in
principle test these predictions by measuring the field--dependence
of width of the Bragg peaks.

Next we come back to our original model (\ref{eq:H}), (\ref{eq:Himp})
keeping the lattice potential $v$ finite. Neglecting the non--Gaussian 
character of
$\phi({\bf x})$, which is justified for $\epsilon\ll 1$,
lowest order perturbation theory in $v$ yields an effective Hamiltonian 
with a mass
   \begin{equation}
   g^2p^2v\,<{\cos{p\phi}}>\sim e^{-p^2{<\phi^2>}/2}\sim
   \left(\frac{L}{L_{\xi}}\right)^{-p^2\bar\eta/2}\,.
   \label{eq:gpv}
   \end{equation}
Comparing this power of $L$ with the $L^{-2}$ behaviour of elastic term,
we conclude, that the periodic perturbation is always relevant if
   $p<p_c=\frac{6}{\pi\sqrt{\epsilon}}$.
In this case, even an arbitrarily weak periodic potential will be relevant
and we regain true translational LRO of the C-phase.

For $p>p_c$, on the other hand, weak periodic pinning is irrelevant, 
but for $p$ close to 
$p_c$ we expect a transition to a commensurate phase for
sufficiently strong $v$. A simple estimate for the threshold value
$v_c$ follows from a comparison of the forces resulting from
impurity and lattice pinning. With $f_{\rm imp}\approx \gamma L^{-2}_{\xi}$
and $f_v\approx\gamma g^2pv$ we get for the transition line 
$v_c=v_c(\Delta)\approx 1/(g^2pL^{2}_{\xi})$ 
(or, inversely, $\Delta_c=\Delta_c(v)\approx (g^2pv)^{(4-D)/2}$ ).

A more accurate description of the transition can be obtained from a
functional renormalization group treatment [10],
which confirms this estimate. The transition turns out to be 
second order with a divergent correlation length $\xi\sim (v-v_c)^{-\nu}$
in the C-phase, here $\nu^{-1}=4\left(\frac{p^2}{p^2_c}-1\right)$.
Moreover, the order parameter for translational LRO
   \begin{equation}
   {\big<\psi\big>}={\big<e^{-i\phi}\big>}\sim
   (v-v_c)^{\beta}\,,\quad \beta =\nu\frac{\pi^2}{18}\epsilon
   \label{eq:psi}
   \end{equation}
is finite in this phase. On the contrary, in the I--phase the quasi--long range
order of (\ref{eq:K1}) is regained. The fixed point, which describes
this transition is at {\it zero temperature}. The temperature eigenvalue 
$-\theta =-2+\epsilon$ appears in the modified hyperscaling relation
$\nu(D-\theta)=2-\alpha$ typical for zero-temperature fixed points..

   \begin{figure}[hbt]
   \centerline{\epsfxsize=7cm
   \epsfbox{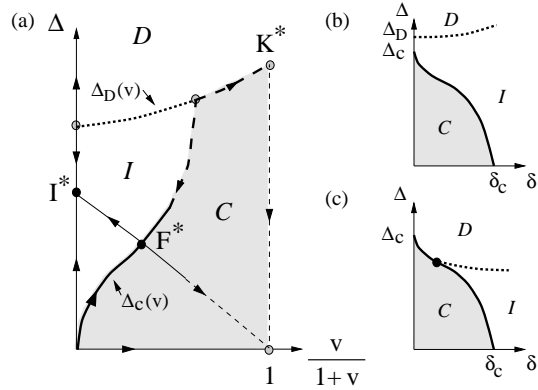}}
   \caption{(a) Schematic RG-flow for $\delta=0$ and $p>p_c$ in the $v-\Delta$
plane. $I$,$C$ and $D$ denote  
the I-, C- and the disordered phase. (b)Possible phase diagram for finite 
$\delta$ at fixed $v$ for $p>p_c$ and (c) for $p<p_c$}
   \end{figure}

As a site remark we note, that these exponents describe for $p=2$
also the transition between the low temperature phase of the {\it random
field Ising model} and the quasi--long range ordered phase of the {\it random 
field $XY$--model}.

Apart from the change in the correlation functions at the
transition there is also a change in the response on a small external
drive $f_{\rm ex}\ll f_{\rm imp},f_v$. The creep velocity $u_{creep}$
can be written in the form
   \begin{equation}
   u_{creep}(f_{\rm ex})\sim e^{-\frac{E_c}{T}
   \left(\frac{f_c}{f_{\rm ex}}\right)^{\mu}}\,.
   \label{eq:u}
   \end{equation}
Here $E_c\approx\gamma\xi^{-2},\;f_c\approx {\rm max}\,
(f_{\rm imp},\gamma\xi^{-2})$ and $\mu=(D-2)/2$ and $\mu =D-1$ for
the I-- and the C--phase, respectively. Because of the
pronounced difference in the creep exponent $\mu$ in both phases
measurement of the creep should give a clear indication about
which phase is present. 

Recently measured I-V curves of the conductor o-TaS$_3$ at
temperatures below 1K can be fitted by (10) with $\kappa=1.5$
-- $2$ [11].  The experimentally observed tendency to larger
$\kappa$ for purer crystals confirms the above interpretation.  In
several materials, such as ${\rm K}_{0.3}{\rm MoO}_3$, the periods are
near $p=4$ commensurability at low temperatures. For this material one
obtains $\xi_0 \approx 10^{-6}$cm [10]. The typical
parameters $\gamma V_0\approx 10^{-2}$eV, $\rho_1=10^{-2}$ and $v_{\rm
  F}=10^7{\rm cm}/{\rm sec}$ yield, after proper rescaling of
anisotropy, the estimate $L_\xi\approx 10^{-4}$cm for an impurity
density of $100$ppm.  Thus it should in principle be possible to see
commensurability effects if the misfit $\delta$ becomes small enough,
i.e. at low temperatures.

\bigskip

So far, we have excluded {\it topological defects}. These can be considered
if we treat $\phi({\bf x})$ as a multivalued field which may jump by
multiples of $2\pi$ along certain surfaces. These surfaces are bounded
by vortex lines. For $\delta=0=v$ it has been shown recently that for
weak enough disorder strength $\Delta<\Delta_D$, the system is stable
with respect to the formation of vortices [12].
However, vortex lines will proliferate for $\Delta>\Delta_D$. At
present it is not clear whether the corresponding transition is
continuous or first order. For $\delta=0$ but $v>0$ we expect that
this transition extends to a line $\Delta_D(v)$ until $v$ reaches a
critical value $v_D$ with $\Delta_D(v_D)=\Delta_c(v_D)$ (see Figure 1). 
For larger $v$ the transition is probably in the
universality class of the $p$-state clock model in a random field,
which has an upper critical dimension $D_c=6$  . A non-zero value of
$\delta$ will in general increase the size of the incommensurate phase,
as schematically scetched in Figure 1.

Our results can also be applied to the pinning of flux line lattices
 in type-II superconductors. In layered superconductors, the
CuO$_2$ planes provide a strong pinning potential favoring, for flux
lines oriented {\it parallel} to the layers, a smectic phase with
translational order present only along the layering axis [13].
The influence of disorder on this phase is described by the CDW model
studied in this paper, if the CDW phase $\phi({\bf x})$ is identified
with the deviations of the smectic layers from their locked-in state.
Also a FLL oriented {\it perpendicular} to the layers in general feels
a weak periodic potential of the underlying crystal, but now the flux
line displacements are described by a vector field. Since also in this
case weak disorder leads only to logarithmically growing transverse
displacement of the flux lines [9], a disorder driven
roughening transition of the CDW type studied above can be expected
for the flux line lattice.

%{\bf 2. SECOND SECTION}\\

%bla bla bla bla bla bla bla bla bla bla bla bla bla bla bla bla bla bla\\

%{\bf 2.1 First Subsection}\\

%bla bla bla bla bla bla bla bla bla bla bla bla bla bla bla bla bla bla\\

%{\bf 2.2 Second Subsection}\\

%bla bla bla bla bla bla bla bla bla bla bla bla bla bla bla bla bla bla\\

%{\bf 3. THIRD SECTION}\\

%bla bla bla bla bla bla bla bla bla bla bla bla bla bla bla bla bla bla\\

%{\bf Acknowledgement}\\

%The authors gratefully acknowledge financial support by the XYZ foundation.\\
\newpage
{\bf References}\\ \nopagebreak

\begin{tabular}{rl}

1. & G.Gr\"uner, {\em Density waves in solids}, (Addison--Wesley, Reading, 1994), \\
   & P. Bak, {\it Rep. Prog. Phys.} {\bf 45}, 587 (1982), \\
   & V. L. Pokrovsky and A. L. Talapov, {\it Theory of Incommensurate
Crystals}, \\ &  Harwood Academic Publishers. 1984.\\  
   & P.M. Chaikin and T.C. Lubensky,{\it Principles of 
Condensed Matter Physics}, Cambridge UP 1995       \\

2. & A.D. Bruce, R.A. Cowley and A.F. Murray, {\it J. Phys. C}{\bf 11}, 351 (1978),\\
3. & S.M. Battacharjee, T. Nattermann and C. Ronnewinkel, {\it Phys.Rev.B} {\bf 58}, 2658 (1998).\\ 
4. & S.N. Coppersmith et al.,{\it Phys. Rev. Lett.} {\bf 46}, 549 (1981).\\
5. & J.M. Kosterlitz,  {\it J. Phys. C} {\bf 10}, 3753 (1977).\\
6. & A. Aharony and P. Bak,{\it Phys. Rev. B} {\bf 23}, 4770 (1981).\\
7. & A.I. Larkin,{\it Sov. Phys. JETP}, {\bf 31}, 784 (1970),\\
   & H. Fukuyama and P.A. Lee, {\it Phys. Rev. B} {\bf 17}, 535 (1978).\\  
8. & S. E. Korshunov, {\it Phys. Rev. B} {\bf48}, 3969 (1993), \\
   & T. Giamarchi and P. Le Doussal,{\it Phys. Rev. Lett.} {\bf 72}, 1530 (1994).\\
9. & T. Emig, S. Bogner and T. Nattermann,{\it Phys. Rev. Lett.} {\bf 83} (1999),
in press.\\
10.& T. Emig, and T. Nattermann, {\it Phys. Rev. Lett.} {\bf 79}, 5090 (1997).\\
11. & S. V. Zaitsev-Zotov, G. Remenyi and P. Monceau,
{\it Phys. Rev. Lett.} {\bf 78}, 1098 (1997).\\
12. & M. Gingras and D. A. Huse, {\it Phys. Rev. B} {\bf 53}, 15193 (1996)\\
    & J. Kierfeld, T. Nattermann and T. Hwa, 
{\it Phys. Rev. B} {\bf 55}, 626 (1997),\\
    & D. S. Fisher, {\it Phys. Rev. Lett.} {\bf 78}, 1964 (1997).\\
13. & L. Balents and D.R. Nelson, {\it Phys. Rev. B} {\bf 52}, 12951
  (1995).\\

\end{tabular}

\end{document}